\documentclass[pra,twocolumn,superscriptaddress,showpacs]{revtex4}

\usepackage{graphicx}

\begin{document}

\title{Quantum anticentrifugal potential in a bent waveguide}

\author{Rossen Dandoloff}
\affiliation{Laboratoire de Physique Th\'{e}orique et
Mod\'{e}lisation (CNRS-UMR 8089), Universit\'{e} de Cergy-Pontoise,
 F-95302 Cergy-Pontoise, France}
\email{rossen.dandoloff@u-cergy.fr}

\author{Victor Atanasov}
\affiliation{Department of Condensed Matter Physics, Sofia University, 5 boul. J. Boucher, 1164 Sofia, Bulgaria}
\email{vatanaso@gmail.com}

\begin{abstract}
We show the existence of an anticentrifugal force for a quantum particle in a bent waveguide. This counterintuitive force due to dimensionality was shown to exist in a flat $R^2$ space but there it needs
an additional $\delta$-like potential at the origin in order to brake the translational invariance and to exhibit localized states. In the case of the bent waveguide
there is no need of any additional potential since here the boundary conditions break the symmetry. The effect may be observed in interference experiments which are sensitive to the additional phase of the wavefunction gained in the bent regions and can find application in distinguishing between straight and bent geometries.
\end{abstract}

\pacs{03.65.-w, 03.65.Ge, 42.50.-p}

\maketitle

It has been noted that dimensionality plays an important role in quantum mechanics. The two-dimensional Euclidean space takes a special place in the quantum world as e.g.
the fractional quantum staistics appears in 2D as well as the quantum anticentrifugal potential that was first shown to exist in 2D \cite{1,2,3} and later in 3D curved space
\cite{4}. Curved 2D surfaces have also been discussed previously\cite{5,6}. The centrifugal potential corresponding to vanishing angular momentum states is attractive rather than repulsive which defies classical intuition. This quantum anticentrifugal force is part of the so called
quantum fictitious forces that appear in two and three space dimensions \cite{7}. This phenomenon is due on one hand to the nonvanishing commutator of the radial momentum $p_r$ and the unit vector in radial direction $\frac{\vec r}{r}$ and on the other hand to the renormalization of the wave function 
so that the Schr\"odinger equation is covariant in the curvilinear coordinates.

Let us now consider a rectangular waveguide with an edge $a$ which is bent along a semicircle with radius $R$ in the $(x,y)$ plane. The semicircle with radius $R$ represents the axis of the rectangular waveguide. The radii of the inner and outer walls of the waveguide are $R-a/2$ and $R+a/2$ respectively. In the Cartesian $(x,y,z)$ coordinate system the "lower" and "upper" walls of the waveguide are situated at $z=0$ and $z=a$ respectively.

\begin{figure}[b]
\begin{center}
\includegraphics[scale=0.25]{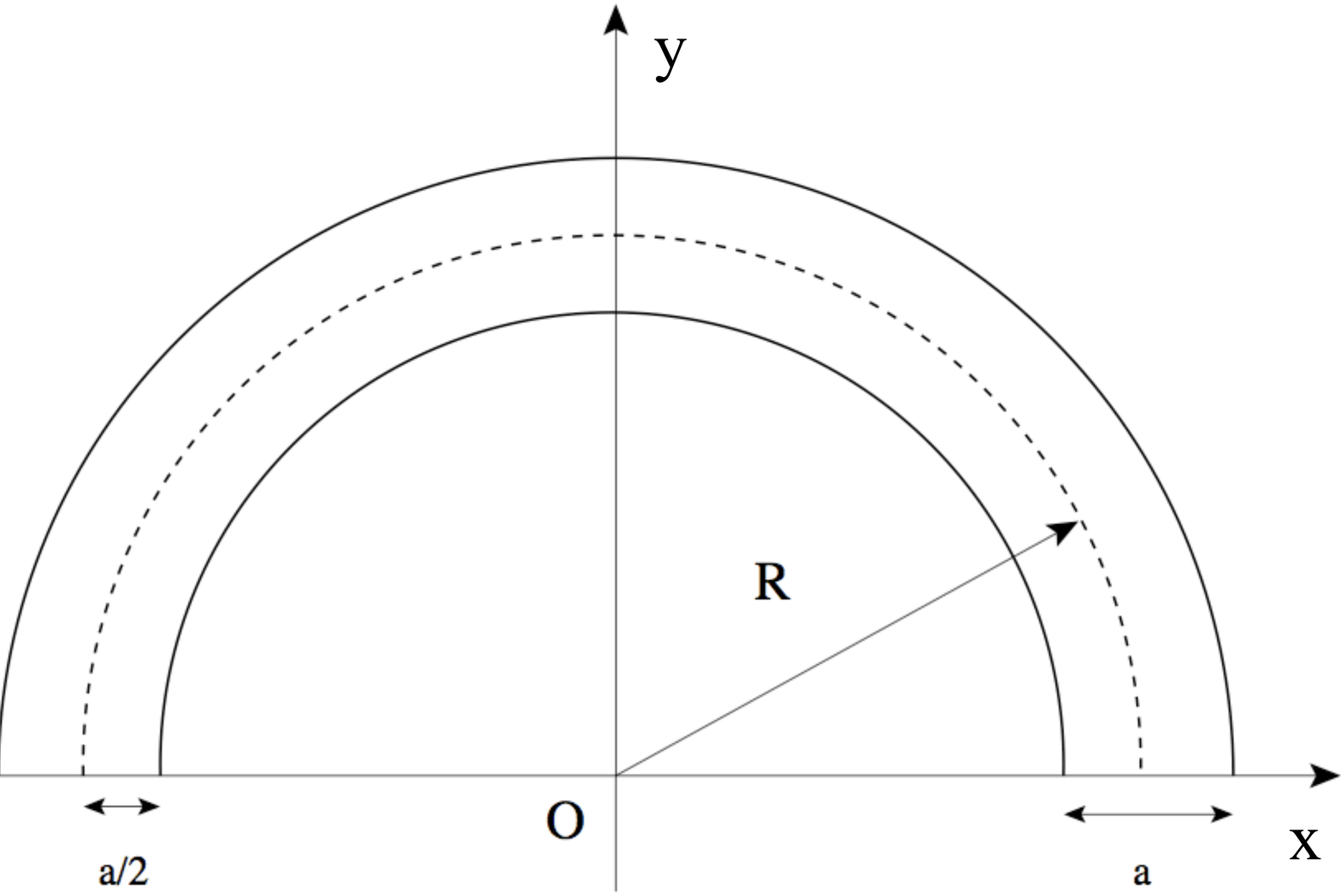}
\caption{\label{waveguidegeom}A projection of the geometry of a bent waveguide onto the $(x,y)$ plane.}
\end{center}
\end{figure}

Now we will introduce a local coordinate system in the bent waveguide. It is depicted in Fig. \ref{waveguidegeom}. We will work with the projection of the rectangular waveguide on the $(x,y)$ plane. This projection
is along the $OZ$-axis. The projection of the central axis of the waveguide on the $(x,y)$-plane is a circle of radius $R.$ The  coordinate $s,$ which is the arc-length of the semi-circle with radius $R,$  is measured from the point $(-R,0).$ The coordinate along the normal to
the circle within the (x,y)-plane is $\xi \in [-\frac{a}{2}, \frac{a}{2}]$ and the coordinate along the $OZ$-axis is $z$. The line element in these coordinates is given by the following expression:

\begin{equation}
dr^2=d\xi^2+dz^2+(1-\kappa \xi)^2ds^2
\end{equation}
where the curvature $\kappa=\frac{1}{R}$ is constant and the corresponding Lam\'e coefficients are $h_z=1$,  $h_\xi=1$,  and   $h_s=(1-\kappa \xi)$. 

Now we are ready to write the Schr\"odinger equation in this coordinate system. The
Laplace operator in the Schr\"odinger equation for the wave function $\Psi$,  has the following form:

\begin{equation}
\Delta\Psi=\frac{\partial^2\Psi}{\partial z^2}+\frac{\partial^2\Psi}{\partial\xi^2}-\frac{k}{(1-\kappa \xi)}\frac{\partial\Psi}{\partial\xi}+\frac{1}{(1-\kappa \xi)^2}\frac{\partial^2\Psi}{\partial s^2}.
\end{equation}

In the curvilinear coordinates the wavefunction acquires the form
\begin{equation}
\Psi=\frac{1}{\sqrt{C}}\frac{\Phi}{\sqrt{1-\kappa \xi} }
\end{equation}
In terms of $\Phi$ the norm of the wavefunction inside the waveguide equals\cite{8}: $\int_{0}^a \int_{-a/2}^{a/2} \int_{0}^{\pi R} |\Phi|^2dzd\xi ds=C,$ where $C=C(a, R).$ In this way one can read off the effective potential from the Schr\"odinger equation in the coordinate system $(\xi, s, z).$ The Laplacian $\Delta\Psi$ becomes:
\begin{eqnarray}
\nonumber \Delta\Psi=&&\frac{1}{(1-\kappa \xi)^{\frac{1}{2}}}\frac{\partial^2\Phi}{\partial\xi^2}+\frac{k^2}{4}\frac{\Phi}{(1-\kappa \xi)^\frac{5}{2}}\\
&&+\frac{\partial^2\Phi}{\partial z^2}+
\frac{1}{(1-\kappa \xi)^2}\frac{\partial^2\Phi}{\partial s^2}
\end{eqnarray}
and the time independent Schrodinger equation $-\frac{\hbar^2}{2M}\Delta\Psi=E\Psi$, (where $M$ is the mass of the particle) takes the following form:
\begin{eqnarray}\label{eq:5}
\nonumber  E\frac{\Phi}{(1-\kappa \xi)^\frac{1}{2}} &=& \frac{\hbar^2}{2M}\left[\frac{1}{(1-\kappa \xi)^{\frac{1}{2}}}\frac{\partial^2\Phi}{\partial\xi^2}+\frac{k^2}{4}\frac{\Phi}{(1-\kappa \xi)^\frac{5}{2}}\right.\\
&& \left.+\frac{\partial^2\Phi}{\partial z^2}+\frac{1}{(1-\kappa \xi)^2}\frac{\partial^2\Phi}{\partial s^2} \right]. 
\end{eqnarray}

The boundary conditions  require that
the wave function becomes zero  on the walls of the wave-guide i.e. at $z=0$ and $z=a.$ (for the $OZ$ direction) and for $\xi=R-a/2$ and $\xi=R+a/2$ (for the radial direction). The wave-guide is open at both ends and therefore there are no boundary conditions for $s=0$ and $s=\pi R$.  For a similar treatment of a straight wave-guide see ref.[9].

Equation (\ref{eq:5}) admits separation of variables. We use the following ansatz 
\begin{equation}
\Phi(s,\xi, z)=e^{i {\rm m} s \kappa}\sin\left(\frac{n\pi}{a}z\right)\Phi_{\rm m}(\xi),
\end{equation}
where $\rm m$ is an integer  quantum number which is the angular momentum of the quantum particle. Indeed $\rm m $ is an eigenvalue of the z-component of the angular momentum operator (which commutes with the Hamiltonian $H=-\frac{\hbar^2}{2M}\Delta,$ therefore the $s-$component of the wavefunction is the eigenfunction of the angular momentum operator. Note that there are no boundary conditions at $s=0$ and $s=\pi R$ and none of the other boundary conditions depend on $s$, hence  the commutator $[J,H ]=0$ is not altered by the boundary conditions) :

\begin{equation}
J= - \frac{i\hbar}{\kappa} \frac{\partial}{\partial s}
\end{equation} 
with eigenfunctions $e^{i{\rm m}s \kappa}.$
We are especially interested in the case of zero angular momentum ${\rm m}=0$ since in this case a quantum anticentrifugal force appears. Now the differential equation for $\Phi_{\rm 0}$ takes the following form:

\begin{equation}\label{Phi_1}
-\frac{\partial^2\Phi_{\rm 0}}{\partial \xi^2}+\left[  \frac{n^2\pi^2}{a^2}-\frac{\kappa^2}{4(1-\kappa \xi)^2}\right] \Phi_{\rm 0}=\frac{2ME}{\hbar^2}\Phi_{\rm 0}
\end{equation}

The above differential equation is an effective Schrodinger equation for $\Phi_{\rm 0}$ with an effective potential $V_{eff}=\left[  \frac{n^2\pi^2}{a^2}-\frac{\kappa^2}{4(1-\kappa \xi)^2}\right].$ In order to solve (\ref{Phi_1}) we denote
\begin{equation}
\mu=1-\kappa \xi, \quad \epsilon^2=\frac{1}{\kappa^2}\left[\frac{2ME}{\hbar^2} -  \frac{n^2\pi^2}{a^2} \right]
\end{equation}
and simplify
\begin{equation}\label{}
-\frac{\partial^2\Phi_{\rm 0}}{\partial \mu^2}-\frac{1}{4\mu^2}\Phi_{\rm 0}= \epsilon^2 \Phi_{\rm 0}.
\end{equation}
Next we introduce
\begin{equation}
\zeta=\epsilon \mu
\end{equation}
and enter the above equation with the ansatz
\begin{equation}
\Phi_{\rm 0}=\sqrt{\zeta}\phi(\zeta)
\end{equation}
to obtain a zeroth order Bessel equation
\begin{equation}
\phi''+\frac{1}{\zeta}\phi'+\phi(\zeta)=0
\end{equation}
which possess oscillating solutions given by $J_0(\zeta)=\frac{1}{2\pi}\int_{-\pi}^{\pi} e^{i \zeta \sin\tau} d \tau.$
In terms of $\xi$ the solutions up to a normalization factor are
\begin{eqnarray}
\nonumber \Phi_{\rm 0}(\xi)&=&\sqrt{  \frac{1}{|k|}\sqrt{\frac{2ME}{\hbar^2} -  \frac{n^2\pi^2}{a^2} } (1-\kappa \xi) } \\
&&J_0 \left( \frac{1}{|k|}\sqrt{\frac{2ME}{\hbar^2} -  \frac{n^2\pi^2}{a^2} } (1-\kappa \xi) \right). \qquad
\end{eqnarray}
Now we need to impose the boundary conditions which will determine the energy eigenvalues. For this purpose we would need the zeroes of the zeroth order Bessel functrion since the wavefunction vanishes at the borders of the waveguide.

\begin{tabular}{|c|c|c|c|c|c|}\hline $\zeta_l: J_0(\zeta_l)=0$ & 2.4048 & 5.5201 & 8.6537 & 11.7915 & 14.9309 \\\hline $l$ & 1 & 2 & 3 & 4 & 5 \\\hline \end{tabular}

Consequently 
\begin{eqnarray}
R\sqrt{\frac{2ME}{\hbar^2} -  \frac{n^2\pi^2}{a^2} } \left[1-\frac{\xi_0}{R}\right]=\zeta_{l}\\
R\sqrt{\frac{2ME}{\hbar^2} -  \frac{n^2\pi^2}{a^2} } \left[1+\frac{\xi_0}{R}\right]=\zeta_{l+w}
\end{eqnarray}
after subtracting the second from the first relation we obtain for the energy
\begin{equation}
E = \frac{\hbar^2}{2M}\left[ \frac{\left( \zeta_{l+w} - \zeta_{l}  \right)^2}{4\xi_0^2} + \frac{n^2\pi^2}{a^2} \right].
\end{equation}
Note, for very large energies $l >> 1$ the asymptotic for the difference of the zeros of the Bessel function is $\zeta_{l+w} - \zeta_{l} \to w \pi.$ For $ l \sim 1,$ we have $\zeta_{l+w} - \zeta_{l} < w \pi$ which reflects the anticentrifugal phenomenon affecting the distribution of the zeroes\cite{1}. 

It is possible to calculate the  Bohm potential corresponding to the bent waveguide 
\begin{equation}
Q=-\frac{\hbar^2}{2M} \frac{\Delta R}{R}.
\end{equation}
We obtain in the $\xi$ coordinate
\begin{equation}
Q(\xi)=\frac{\hbar^2}{2M}  \frac{\kappa^2 \left( a_2 \xi^2 + a_1 \xi + a_3 \right)}{4 \xi_0^2 (1-\kappa \xi)^2},
\end{equation}
where
\begin{eqnarray}
a_1&=& -2 \frac{(\zeta_{l+w}-\zeta_l)^2}{\kappa} \\
a_2&=& (\zeta_{l+w}-\zeta_l)^2   \\\label{a3}
a_3&=& \frac{(\zeta_{l+w}-\zeta_l)^2}{\kappa^2}+\xi_0^2  
\end{eqnarray}
The dependence of the Bohm potential on $\xi$ is very weak and we can approximate it with an effective one dimensional rectangular barrier which corresponds to the case of a very thin waveguide. The height of this barrier is the value of $Q$ at $\xi=0.$ Adding the contribution form the $z$ coordinate we write the total Bohm potential
\begin{equation}
Q_0=\frac{\hbar^2}{2M} \left[  \frac{(\zeta_{l+w}-\zeta_l)^2}{4\xi_0^2} +\frac{\kappa^2}{4} + \frac{n^2\pi^2}{a^2}  \right].
\end{equation}

With the help of the Bohm potential we can translate the effect of the boundary conditions in terms of an additional potential affecting the quantum motion. It changes the interference picture according to
\begin{eqnarray}
p'&&=\sqrt{2M(E-Q_0)}\\
\nonumber &&\approx p - \frac{\hbar^2}{p} \left[  \frac{(\zeta_{l+w}-\zeta_l)^2}{4\xi_0^2} +\frac{\kappa^2}{4} + \frac{n^2\pi^2}{a^2}  \right]
\end{eqnarray}
which yields for the phase shift of the wavefunction the following
\begin{equation}
\Delta \phi = l \Delta p = \frac{\pi}{\kappa}\Delta p.
\end{equation}
This phase shift is minimized for $n=0$ and similarly for the transverse states in $\xi$
\begin{equation}
\Delta \phi _{min} = \frac{\pi \hbar^2 \kappa}{4 p}.
\end{equation} 
Inroducing the de Broglie wavelength $\lambda=h/p$ the above reduces to
\begin{equation}
\Delta \phi _{min} = \frac{\hbar \lambda \kappa}{8}.
\end{equation} 
This is an irreducible amount which is always present when the geometry of the waveguide is curved and can be used in an interference experiment to distinguish between curved and straight geometries.

The quantum anticentrifugal force on the central line defined as the mean value of the gradient of the effective potential is given by the following expression:
\begin{equation}
F=-\frac{\hbar^2}{2M}\frac{dV_{eff}}{d\xi}_{|_{\xi=0}}=\frac{\hbar^2}{2M}\frac{\kappa^3}{2}=\frac{\hbar^2}{2M}\frac{1}{2R^3} 
 \end{equation}
 where $R$ is the radius of the waveguide. We note the unusual one over $(distance)^3$ dependence of this anticentrifugal force for ${\rm m}=0$. This defies classical intuition.

In conclusion, it should be noted the similarity of the stationary Sch\"odinger equation and the Helmhotz equation for the TE  modes in a 
waveguide. Consequently, in the interference picture of an electromagnetic process in a bent waveguide, one should find a similar pattern of interference as in the quantum case.


\begin{thebibliography}{77}

\bibitem{1} M.A. Cirone, K. Rzkazewski, W.P. Schleich, F. Straub and J.A. Wheeler, Phys. Rev. A, {\bf 65}, 022101-1, (2001)

\bibitem{2} I. Bialynicki-Birula, M.A. Cirone, J.P. Dahl, M. Fedorov and W.P. Schleich, Phys. Rev. Lett., {\bf 89}, 060404-1, (2002)

\bibitem{3} W.P. Schleich and J.P. Dahl, Phys. Rev. A, {\bf 65}, 052109, (2002); J.P. Dahl and W.P. Schleich, Phys. Rev. A, {\bf 65}, 022109, (2002).

\bibitem{4} R. Dandoloff, Phys.Lett. A,  {\bf 373}, (2009), 2667.

\bibitem{5} R. Dandoloff, A. Saxena and B. Jensen, Phys. Rev. {\bf A}, {\bf 81}, 014102 (2010).

\bibitem{6} V. Atanasov and R. Dandoloff, Phys.Lett. {\bf A} {\bf 371}, 118 (2007);  Phys.Lett. {\bf A} {\bf 372}, 6141 (2008).

\bibitem{7} I. Bialynicki-Birula, M.A. Cirone, J.P. Dahl, T.H. Seligman, F. Straub and W.P. Schleich, Fortschr.Phys., {\bf 50}, 599, (2002).

\bibitem{8} R.C.T. da Costa, Phys. Rev.  A {\bf 23}, 1982 (1981); J. Goldstone and R.L. Jaffe, Phys.Rev. B {\bf 45}, 14100 (1992).

\bibitem{9} J.-M. Levy-Leblond, Phys. Lett.  {\bf A} {\bf 125},  (1987), 441.

\end{thebibliography}
\end{document}